\documentclass[a4paper,11pt]{article}
\usepackage{amssymb}
\usepackage[dvips]{graphicx}
\usepackage[usenames]{color}
\usepackage{amsmath}
\usepackage{fullpage}
\usepackage{boxedminipage}
\usepackage{listings}
\usepackage{minitoc}
\usepackage{wrapfig}
\usepackage{latexsym}

\makeatletter \@addtoreset{equation}{section} \makeatother

\begin{document}

\renewcommand{\[}{\begin{equation}} \renewcommand{\]}{\end{equation}} %
\renewcommand{\>}{\rangle}

\begin{titlepage}

    \thispagestyle{empty}
    \begin{flushright}
        \hfill{CERN-PH-TH/2011-038}\\
    \end{flushright}

    \vspace{15pt}
    \begin{center}
        { \huge{\textbf{Perturbative and Non-Perturbative Aspects\\\vspace{10pt}of $\mathcal{N}=8$ Supergravity}}}\vspace{25pt}

        \vspace{55pt}

         {\large{\bf Sergio Ferrara$^{1,2}$ and\ Alessio Marrani$^{1}$}}

        \vspace{15pt}

        {$1$ \it Physics Department,Theory Unit, CERN, \\
        CH 1211, Geneva 23, Switzerland\\
        \texttt{sergio.ferrara@cern.ch}\\
        \texttt{Alessio.Marrani@cern.ch}}

        \vspace{10pt}

        {$2$ \it INFN - Laboratori Nazionali di Frascati, \\
        Via Enrico Fermi 40,00044 Frascati, Italy}

 \vspace{50pt}

        \noindent \textit{Contribution to the Proceedings of the International School of Subnuclear
        Physics,\\\vspace{2pt}48th Course: ``What is Known and Unexpected at LHC",\\\vspace{2pt}Erice, Italy, 29 August -- 7 September 2010,\\\vspace{2pt}Based on Lectures given by S. Ferrara}
\end{center}

\vspace{50pt}

\begin{abstract}
Some aspects of quantum properties of $\mathcal{N}=8$ supergravity
in four dimensions are discussed for non-practitioners.

At perturbative level, they include the Weyl trace anomaly as well
as composite duality anomalies, the latter being relevant for
perturbative finiteness. At non-perturbative level, we briefly
review some facts about extremal black holes, their
Bekenstein-Hawking entropy and attractor flows for single- and two-
centered solutions.
\end{abstract}

\end{titlepage}
\newpage

\section{Lecture I\newline
On ``Quantum'' $\mathcal{N}=8$, $d=4$ Supergravity}

$\mathcal{N}=8$, $d=4$ ``quantum'' supergravity may be defined by starting
with the Einstein-Hilbert action, and setting ``perturbative'' Feynman rules
as a \textit{bona fide} gauge theory of gauge particles of spin $2$, the
gravitons. In supersymmetric gravity theories with $\mathcal{N}$-extended
supersymmetry in $d=4$ space-time dimensions, the massless particle content
is given by
\begin{equation}
\binom{\mathcal{N}}{k}\equiv \frac{\mathcal{N}!}{k!\left( \mathcal{N}%
-k\right) !}\text{ particles~of~\textit{helicity}~}\lambda =2-\frac{k}{2},
\end{equation}
where $k_{\max }=\mathcal{N}$, and $\mathcal{N}\leqslant 8$ if $\left|
\lambda \right| \leqslant 2$ is requested (namely, no higher spin fields in
the massless spectrum).

One possible approach to ``quantum'' supergravity is to consider it as it
comes from $M$-theory restricted to the massless sector. The problem is that
this theory, even if preserving maximal $\mathcal{N}=8$ supersymmetry in $%
d=4 $ space time dimensions (corresponding to $32=8\times 4$
supersymmetries), is \textit{not} uniquely defined, because of the multiple
choice of internal compactification manifolds and corresponding duality
relations:
\begin{equation}
\begin{array}{ll}
\begin{array}{l}
\mathbf{I}.~M_{11}\longrightarrow M_{4}\times T_{7} \\
~
\end{array}
&
\begin{array}{l}
\text{(}GL^{+}(7,\mathbb{R})\text{ and }SO\left( 7\right) ~\text{manifest);}
\\
~
\end{array}
\\
\begin{array}{l}
\mathbf{II}.~M_{11}\longrightarrow AdS_{4}\times S^{7} \\
~
\end{array}
&
\begin{array}{l}
\text{(}SO\left( 8\right) ~\text{manifest,~\textit{gauged});} \\
~
\end{array}
\\
\mathbf{III}.~M_{11}\longrightarrow M_{4}\times T_{7,\mathcal{R}} & \text{(}%
SL(8,\mathbb{R})\text{ and }SO\left( 8\right) ~\text{manifest),}
\end{array}
\label{various-N=8,d=4}
\end{equation}
where $T_{7}$ is the $7$-torus and $S^{7}$ is the $7$-sphere. $T_{7,\mathcal{%
R}}$ denotes the case in which, according to Cremmer and Julia \cite{CJ},
the dualization of $21$ vectors and $7$ two-forms makes $SL(8,\mathbb{R})$
(in which $GL^{+}(7,\mathbb{R})$ is maximally embedded) manifest as maximal
non-compact symmetry of the Lagrangian. Note that in case $\mathbf{III}$ one
can further make $E_{7\left( 7\right) }$ (and its maximal compact subgroup $%
SU\left( 8\right) $) manifest \textit{on-shell}, by exploiting a Cayley
transformation supplemented by a rotation through $SO\left( 8\right) $ gamma
matrices on the vector\ $2$-form field strengths \cite{CJ,HW}. As we discuss
further below, $E_{7\left( 7\right) }$ can be promoted to a Lagrangian
symmetry if one gives up manifest diffeomorphism invariance, as given by
treatment in \cite{Hillmann}, then used in the anomaly study of \cite{BHN}.

It is worth remarking that $\mathcal{N}=8$, $d=4$ \textit{gauged}
supergravity with gauge group $SO(8)$ cannot be used for electroweak and
strong interactions model building, because
\begin{equation}
SO\left( 8\right) \nsupseteq SU\left( 3\right) \times SU\left( 2\right)
\times U\left( 1\right) .
\end{equation}
Furthermore, also the cosmological term problem arises out: the vacuum
energy in anti De Sitter space $AdS_{4}$ is much higher than the vacuum
energy in Standard Model of non-gravitational interactions (see \textit{e.g.}
the discussion in \cite{Slansky}). However, by exploiting the $%
AdS_{4}/CFT_{3}$ correspondence, theory $\mathbf{II}$ of (\ref
{various-N=8,d=4}) recently found application in $d=3$ condensed matter
physics (see \textit{e.g.} \cite{Hartnoll} for a review and list of Refs.).
Furthermore, the recently established fluid-gravity correspondence was
object of many studies (see \textit{e.g.} \cite{Rangamani} for recent
reviews and lists of Refs.).

The fundamental massless fields (and the related number $\sharp $ of degrees
of freedom) of $M$-theory in $d=11$ flat space-time dimensions are
\begin{equation}
\begin{array}{lll}
\begin{array}{l}
g_{\mu \nu }~\text{(\textit{graviton})}: \\
~
\end{array}
&
\begin{array}{l}
\sharp =\frac{\left( d-1\right) \left( d-2\right) }{2}-1, \\
~
\end{array}
&
\begin{array}{l}
\text{in~}d=11:\sharp =44; \\
~
\end{array}
\\
\begin{array}{l}
\Psi _{\mu \alpha }~\text{(\textit{gravitino})}: \\
~
\end{array}
&
\begin{array}{l}
\sharp =(d-3)2^{(d-3)/2}, \\
~
\end{array}
&
\begin{array}{l}
\text{in~}d=11:\sharp =128; \\
~
\end{array}
\\
A_{\mu \nu \rho }~\text{(\textit{three-form})}: & \sharp =\frac{\left(
d-2\right) \left( d-3\right) \left( d-4\right) }{3!}, & \text{in~}%
d=11:\sharp =84.
\end{array}
\end{equation}
Because a $\left( p+1\right) $-form (``Maxwell-like'' gauge field) $A_{p+1}$
couples to $p$-dimensional extended objects, and its ``magnetic'' dual $%
B_{d-p-3}$ couples to $\left( d-p-4\right) $-dimensional extended objects,
it follows that the fundamental (massive) objects acting as sources of the
theory are $M2$- and $M5$-branes.

In general, a compactification on an $n$-torus $T_{n}$ has maximal manifest
non-compact symmetry $GL^{+}\left( n,\mathbb{R}\right) \sim \mathbb{R}%
^{+}\times SL\left( n,\mathbb{R}\right) $. The metric $g_{IJ}$ of $T_{n}$
parametrizes the $n\left( n+1\right) /2$-dimensional coset $\mathbb{R}%
^{+}\times \frac{SL\left( n,\mathbb{R}\right) }{SO\left( n\right) }$,
whereas the Kaluza-Klein vectors $g_{\widehat{\mu }}^{I}$ are in the $%
\mathbf{n}^{\prime }$ irrep. of $GL^{+}\left( n,\mathbb{R}\right) $ itself.
By reducing $M$-theory on $T_{7}$ a $d=4$ theory with maximal ($\mathcal{N}%
=8 $) local supersymmetry arises. By splitting the $d=11$ space-time index $%
\mu =0,1,...,10$ as $\mu =\left( \widehat{\mu },I\right) $, where $\widehat{%
\mu }=0,1,...3$ is the $d=4$ space-time index, and $I=1,...,7$ is the
internal manifold index, the bosonic degrees of freedom of $M$-theory split
as follows (below (\ref{c}), for simplicity's sake we will then refrain from
hatting the $d=4$ curved indices):
\begin{eqnarray}
g_{\mu \nu } &\longrightarrow &\left\{
\begin{array}{ll}
\begin{array}{l}
g_{\widehat{\mu }\widehat{\nu }}~\text{(}d=4~\text{\textit{graviton})}, \\
~
\end{array}
&
\begin{array}{l}
\mathbf{1}+\mathbf{1}; \\
~
\end{array}
\\
\begin{array}{l}
g_{\widehat{\mu }}^{I}~\text{(\textit{vectors})}, \\
~
\end{array}
&
\begin{array}{l}
\mathbf{7}^{\prime }; \\
~
\end{array}
\\
g_{IJ}~\text{(\textit{scalars})}, & \mathbf{28};
\end{array}
\right. \\
&&  \notag \\
A_{\mu \nu \rho } &\longrightarrow &\left\{
\begin{array}{ll}
\begin{array}{l}
A_{\widehat{\mu }\widehat{\nu }\widehat{\rho }}~\text{(}d=4~\text{\textit{%
domain~wall})}, \\
~
\end{array}
&
\begin{array}{l}
\sharp =0; \\
~
\end{array}
\\
\begin{array}{l}
A_{\widehat{\mu }\widehat{\nu }I}~\text{(\textit{antisymmetric tensors :
strings})}, \\
~
\end{array}
&
\begin{array}{l}
\mathbf{7}; \\
~
\end{array}
\\
\begin{array}{l}
A_{\widehat{\mu }IJ}~\text{(\textit{vectors})}, \\
~
\end{array}
&
\begin{array}{l}
\mathbf{21}; \\
~
\end{array}
\\
A_{IJK}~\text{(\textit{scalars})}, & \mathbf{35},
\end{array}
\right.  \label{c}
\end{eqnarray}
where the indicated irreps. pertain to the maximal manifest non-compact
symmetry $GL^{+}(7,\mathbb{R})$, whose maximal compact subgroup is $SO\left(
7\right) $. The $28$ scalars $g_{IJ}$ (metric of $T_{7}$) parametrize the
coset $\mathbb{R}^{+}\times \frac{SL\left( 7,\mathbb{R}\right) }{SO\left(
7\right) }$.

By switching to formulation $\mathbf{III}$ of (\ref{various-N=8,d=4}) \cite
{CJ}, the $7$ antisymmetric rank-$2$ tensors $A_{\widehat{\mu }\widehat{\nu }%
I}$ (sitting in the $\mathbf{7}$ of $GL^{+}\left( 7,\mathbb{R}\right) $ can
be dualized to scalars $\phi ^{I}$ (in the $\mathbf{7}^{\prime }$ of $%
GL^{+}\left( 7,\mathbb{R}\right) $), and therefore one obtains $35+28+7=70$
scalar fields. It is worth remarking that in Cremmer and Julia's \cite{CJ}
theory the gravitinos $\psi _{I}$ and the gauginos $\chi _{IJK}$
respectively have the following group theoretical assignment\footnote{%
As evident from (\ref{fermions-III}), we use a different convention with
respect to \cite{Slansky} (see \textit{e.g.} Table 36 therein). Indeed, we
denote as $\mathbf{8}_{v}$ of $SO\left( 8\right) $ the irrep. which
decomposes into $\mathbf{7}+\mathbf{1}$ of $SO\left( 7\right) $, whereas the
two spinorial irreps. $\mathbf{8}_{s}$ and $\mathbf{8}_{c}$ both decompose
into $\mathbf{8}$ of $SO\left( 7\right) $. The same change of notation holds
for $\mathbf{35}$ and $\mathbf{56}$ irreps..} ($I$ in $\mathbf{8}$ of $%
SU\left( 8\right) $):
\begin{equation}
\text{theory }\mathbf{III~}\text{\cite{CJ}}:\left\{
\begin{array}{l}
\psi _{I}:\underset{\mathbf{8}}{~SO\left( 7\right) }\subset \underset{%
\mathbf{8}_{s}}{SO\left( 8\right) }\subset \underset{\mathbf{8}}{SU\left(
8\right) }; \\
\\
\chi _{IJK}:\underset{\mathbf{8}+\mathbf{48}}{~SO\left( 7\right) }\subset
\underset{\mathbf{56}_{s}}{SO\left( 8\right) }\subset \underset{\mathbf{56}}{%
SU\left( 8\right) }.
\end{array}
\right.  \label{fermions-III}
\end{equation}
Thus, in this theory the $70$ scalars arrange as
\begin{equation}
\text{theory }\mathbf{III~}\text{\cite{CJ}}:\underset{\left( \sharp
=70\right) }{s=0\text{~\textit{dofs}}}:~\underset{\mathbf{1}+\mathbf{7}+%
\mathbf{21}+\mathbf{35}}{SO\left( 7\right) }\subset \underset{\mathbf{35}%
_{v}+\mathbf{35}_{c}}{SO\left( 8\right) }\subset \underset{\mathbf{70}}{%
SU\left( 8\right) },  \label{scalars-III}
\end{equation}
where $\mathbf{70}$ is the rank-$4$ completely antisymmetric irrep. of $%
SU\left( 8\right) $, the maximal compact subgroup of the $U$-duality group $%
E_{7\left( 7\right) }$ (also called $\mathcal{R}$-symmetry).

On the other hand, also the vector fields $A_{\widehat{\mu }IJ}$ (sitting in
the $\mathbf{21}$ of $GL^{+}(7,\mathbb{R})$) can be dualized to $A_{\widehat{%
\mu }}^{IJ}$ (sitting in the $\mathbf{21}^{\prime }$ of $GL^{+}(7,\mathbb{R}%
) $). Together with $g_{\widehat{\mu }}^{I}$, the ``electric'' and
``magnetic'' vector degrees of freedom can thus be arranged as follows:
\begin{equation}
\underset{\left( \sharp =56\right) }{s=1\text{~\textit{dofs}}}:~\left\{
\begin{array}{l}
\underset{\mathbf{7}^{\prime }+\mathbf{21}^{\prime }+\mathbf{7}+\mathbf{21}}{%
GL^{+}\left( 7,\mathbb{R}\right) }\subset \underset{\mathbf{28}^{\prime }+%
\mathbf{28}}{SL\left( 8,\mathbb{R}\right) }\subset \underset{\mathbf{56}}{%
E_{7(7)}}; \\
\\
\underset{\mathbf{7}+\mathbf{21}+\mathbf{7}+\mathbf{21}}{SO\left( 7\right) }%
\subset \underset{\mathbf{28}+\mathbf{28}}{SO\left( 8\right) }\subset
\underset{\mathbf{28}+\overline{\mathbf{28}}}{SU\left( 8\right) }.
\end{array}
\right.  \label{vectors}
\end{equation}

The counting of degrees of freedom is completely different in the \textit{%
gauged} maximal supergravity theory $\mathbf{II}$ of (\ref{various-N=8,d=4}%
), based on the $AdS_{4}\times S^{7}$ solution of $d=11$, $\mathcal{N}=1$ $M$%
-theory field equations; in this framework, rather than using torus indices
as in theories $\mathbf{I}$ and $\mathbf{III}$ of (\ref{various-N=8,d=4}),
Killing vector/spinor techniques are used (for a discussion, see \textit{e.g.%
} \cite{BEdWN}, and the lectures \cite{DP}, and Refs. therein). However, the
$70$ scalars still decompose\footnote{%
There are three distinct $35$-dimensional $SO\left( 8\right) $ irreps.,
usually denoted as $\mathbf{35}_{v}$, $\mathbf{35}_{s}$ and $\mathbf{35}_{c%
\text{ }}$, obeying the relations:
\begin{equation*}
\left( ab\right) \leftrightarrow \left[ ABCD\right] _{+};~\left[ abcd\right]
_{+}\leftrightarrow \left[ ABCD\right] _{-};~\left[ abcd\right]
_{-}\leftrightarrow \left( AB\right) ,
\end{equation*}
where $a,b=1,...,8$ are in $\mathbf{8}_{v}$, $A,B,C,D=1,...,8$ are in $%
\mathbf{8}_{s}$ (or in $\mathbf{8}_{c}$), and ``$\pm $'' denotes
self-dual/anti-self-dual irreps.. For a discussion, see \textit{e.g.} \cite
{CJ} and \cite{DP}.} as $\mathbf{35}_{v}+\mathbf{35}_{c}$ of $SO\left(
8\right) $ but, with respect to the chain of branchings (\ref{scalars-III}),
they lack of any $SO\left( 7\right) $ interpretation. It is worth recalling
that a formulation of this theory directly in $d=4$ yields to the de Wit and
Nicolai's $\mathcal{N}=8$, $d=4$ \textit{gauged} supergravity \cite{dWN}.

Since the $70$ scalar fields fit into an unique irrep. of $SU(8)$, it
follows that they parameterize a non-compact coset manifold $\frac{G}{%
SU\left( 8\right) }$. Indeed, the $SU\left( 8\right) $ under which both the
scalar fields and the fermion fields transform is the \textit{``local''} $%
SU\left( 8\right) $, namely the stabilizer of the scalar manifold. On the
other hand, the $SU\left( 8\right) $ appearing in the second line of (\ref
{vectors}), \textit{i.e.} the one under which the vector $2$-form
self-dual/anti-self-dual field strengths transform, is the \textit{``global''%
} $SU\left( 8\right) $ ($\mathcal{R}$-symmetry group). Roughly speaking, the
physically relevant group $SU\left( 8\right) $ is the diagonal one in the
product $SU_{\text{local}}\left( 8\right) \times SU_{\text{global}}\left(
8\right) $ (see also discussion below).

Remarkably, there exists an \textit{unique} simple, non-compact Lie group
with real dimension $70+63=133$ and which embeds $SU\left( 8\right) $ as its
maximal compact subgroup: this is the real, non-compact split form $%
E_{7\left( 7\right) }$ of the exceptional Lie group $E_{7}$, thus giving
rise to the symmetric, rank-$7$ coset space
\begin{equation}
\frac{E_{7\left( 7\right) }}{SU\left( 8\right) /\mathbb{Z}_{2}},
\end{equation}
which is the scalar manifold of $\mathcal{N}=8$, $d=4$ supergravity ($%
\mathbb{Z}_{2}$ is the kernel of the $SU\left( 8\right) $-representations of
even rank; in general, spinors transform according to the double cover of
the stabilizer of the scalar manifold; see \textit{e.g.} \cite
{Yokota,AFZ-rev}).

$E_{7\left( 7\right) }$ acts as electric-magnetic duality symmetry group
\cite{GZ}, and its maximal compact subgroup $SU\left( 8\right) $ has a
chiral action on fermionic as well as on (the vector part of the) bosonic
fields. While the chiral action of $SU\left( 8\right) $ on fermions directly
follows from the chirality (complex nature) of the relevant irreps. of $%
SU\left( 8\right) $ (as given by Eq. (\ref{fermions-III})), the chiral
action on vectors is a crucial consequence of the electric-magnetic duality
in $d=4$ space-time dimensions. Indeed, this latter allows for
``self-dual/anti-self-dual'' complex combinations of the field strengths,
which can then fit into complex irreps. of the stabilizer $H$ of the coset
scalar manifold $G/H$ itself. For the case of maximal $\mathcal{N}=8$
supergravity, the relevant chiral complex irrep. of $H=SU\left( 8\right) $
is the rank-$2$ antisymmetric $\mathbf{28}$, as given by Eq. (\ref{vectors}).

Note that if one restricts to the $SL\left( 8,\mathbb{R}\right) $-covariant
sector, the chirality of the action of electric-magnetic duality is spoiled,
because the maximal compact subgroup of $SL\left( 8,\mathbb{R}\right) $,
namely $SO\left( 8\right) $, has not chiral irreps.

Composite (sigma model $G/H$) anomalies can arise in theories in which $G$
has a maximal compact subgroup with a \textit{chiral} action on bosons
and/or fermions (see \textit{e.g.} \cite{DFG,Marcus,BHN}). Surprising
cancellations among the various contributions to the composite anomaly can
occur as well. An example is provided by $\mathcal{N}=8$, $d=4$ supergravity
itself, in which standard anomaly formul\ae\ yield the remarkable result
\cite{Marcus,BHN}
\begin{equation}
3Tr_{\mathbf{8}}X^{3}-2Tr_{\mathbf{28}}X^{3}+Tr_{\mathbf{56}}X^{3}=\left(
3-8+5\right) Tr_{\mathbf{8}}X^{3}=0,  \label{(caN=8)=0}
\end{equation}
where $X$ is any generator of the Lie algebra $\frak{su}(8)$ of the \textit{%
rigid} (\textit{i.e.} \textit{global)} $SU(8)$ group ($\mathcal{R}$%
-symmetry). In light of the previous considerations, the first and third
contributions to (\ref{(caN=8)=0}) are due to fermions: the $8$ gravitinos $%
\psi _{A}$ and the $56$ spin-$\frac{1}{2}$ fermions $\chi _{ABC}$,
respectively, whereas the second contribution is due to the $28$ chiral
vectors. Note that, for the very same reason, the \textit{local} $SU(8)$
(stabilizer of the non linear sigma-model of scalar fields), under which
only fermions do transform\footnote{%
Also scalar fields transform under \textit{local} $SU\left( 8\right) $, but
they do not contribute to the composite anomaly, because they sit in the
\textit{self-real} (and thus \textit{non-chiral}) rank-$4$ antisymmetric
irrep. $\mathbf{70}$ of $SU\left( 8\right) $.}, would be \textit{anomalous}
\cite{DFG}.\texttt{\ }In an analogous way, in \cite{Marcus} it was
discovered that $\mathcal{N}=6$ and $\mathcal{N}=5$ ``pure'' supergravities
are \textit{composite} \textit{anomaly-free}, whereas $\mathcal{N}\leqslant
4 $ theories are not.

A crucial equivalence holds at the homotopical level:
\begin{equation}
E_{7\left( 7\right) }\cong \left( SU\left( 8\right) /\mathbb{Z}_{2}\right)
\times \mathbb{R}^{70},
\end{equation}
implying that the two group manifolds have the same De Rham cohomology. This
is a key result, recently used in \cite{BHN} to show that the aforementioned
absence of $SU(8)$ current anomalies yield to the absence of anomalies for
the non-linearly realized $E_{7\left( 7\right) }$ symmetry, thus implying
that the $E_{7\left( 7\right) }$ continuous symmetry of classical $\mathcal{N%
}=8$, $d=4$ supergravity is preserved at all orders in perturbation theory
(see \textit{e.g.} \cite{4-loop,K-1,BFK-1,BFK-2,Vanhove,Dixon,K-2,Freedman}%
). This implies the perturbative finiteness of supergravity at least up to
seven loops; Bern, Dixon \textit{et al.} explicitly checked the finiteness
up to four loops included \cite{4-loop} (computations at five loops, which
might be conclusive, are currently in progress; for a recent review, see
\textit{e.g.} \cite{Bern-last}).

In order to achieve the aforementioned result on the anomalies of $%
E_{7\left( 7\right) }$, in \cite{BHN} the manifestly $E_{7\left( 7\right) }$%
-covariant Lagrangian formulation of $\mathcal{N}=8$, $d=4$ supergravity
\cite{Hillmann} was exploited, by using the ADM decomposition of the $d=4$
metric, namely\footnote{%
We use\textit{\ }units in which the Newton gravitational constant $G$ and
the speed of light in vacuum $c$ are all put equal to $1$.}:
\begin{equation}
g_{\mu \nu }dx^{\mu }dx^{\nu }=-N^{2}dt^{2}+h_{ij}\left(
dx^{i}+N^{i}dt\right) \left( dx^{j}+N^{j}dt\right) ,
\end{equation}
with lapse $N$ and shift $N^{i}$ ($h_{ij}$ is the metric on the spatial
slice). Within this approach \cite{Henneaux-T,Hillmann}, the diffeomorphism
symmetry is not realized in the standard way on the vector fields: the $28$
vector fields $A_{\mu }^{\Lambda }$ of the original formulation \cite{CJ,dWN}
are replaced by $56$ vector fields $A_{i}^{\mathbb{B}}$ ($\mathbb{B}%
=1,...,56 $) with only spatial components, which recover the number of
physical degrees of freedom by switching to an Hamiltonian formulation.
Besides the $56\times 56$ symplectic metric $\Omega $:
\begin{equation}
\Omega ^{T}=-\Omega ,~\Omega ^{2}=-\mathbb{I},
\end{equation}
a crucial quantity is the scalar field-dependent $56\times 56$ symmetric
matrix $\mathcal{M}$ (see Eq. (\ref{M-expl}) below), which is symplectic
(see \textit{e.g.} \cite{CDF-rev}):
\begin{equation}
\mathcal{M}\Omega \mathcal{M}=\Omega ,  \label{M}
\end{equation}
and negative definite due to the positivity of the vector kinetic terms (see
also discussion below). $\mathcal{M}$ allows for the introduction of a
symplectic, scalar field-dependent complex structure:
\begin{equation}
\mathcal{J}\equiv \mathcal{M}\Omega \Rightarrow \mathcal{J}^{2}=\mathcal{M}%
\Omega \mathcal{M}\Omega =-\mathbb{I}.  \label{J}
\end{equation}
Thus, the equations of motion of the $56$ vector fields $A_{i}^{\mathbb{B}}$
can be expressed as a twisted self-duality condition\footnote{%
For interesting recent developments on twisted self-duality, see \cite
{Brunster-Henneaux}.} \cite{CJ} for their super-covariant fields strengths $%
\hat{F}_{\mu \nu }^{\mathbb{A}}$, namely (see \cite{Hillmann,BHN} for
further elucidation)
\begin{equation}
\hat{F}_{\mu \nu }^{\mathbb{A}}=-\frac{1}{2\sqrt{\left| g\right| }}\epsilon
_{\mu \nu }^{~~\rho \sigma }\mathcal{J}_{~\mathbb{B}}^{\mathbb{A}}\hat{F}%
_{\mu \nu }^{\mathbb{B}}.
\end{equation}
Although the time components $A_{0}^{\mathbb{B}}$ do not enter the
Lagrangian, they appear when solving the equations of motion for the spatial
components $A_{i}^{\mathbb{B}}$, and diffeomorphism covariance is recovered
on the solutions of the equations of motion \cite{Hillmann,BHN}.

From power counting arguments in quantum gravity, an $n$-loop counterterm
contains $2n+2$ derivatives, arranged such that it does not vanish \textit{%
on-shell}. In $\mathcal{N}=8$ supergravity the first (non-BPS) full
superspace integral which is $E_{7\left( 7\right) }$-invariant is the
\textit{super-Vielbein} superdeterminant, which may contain as last
component a term $\sim \partial ^{8}R^{4}$ (see \textit{e.g.} \cite
{Howe-Lindstrom}, and also \cite{BG}), then possibly contributing to a
divergence in the four-graviton amplitude. However, in \cite{K-2} R. Kallosh
argued that, by exploiting the light-cone formulation, candidate
counterterms can be written in chiral, but \textit{not} in real, light-cone
superspace. This would then imply the ultraviolet finiteness of $\mathcal{N}%
=8$, $d=4$ supergravity, \textit{if} supersymmetry and $E_{7\left( 7\right) }
$ symmetry are non-anomalous. Recently, in \cite{Kallosh-proof} the latter
symmetry was advocated by the same author to imply ultraviolet finiteness of
the theory to all orders in perturbation theory.

A puzzling aspect of these arguments is that string theory certainly
violates continuous $E_{7\left( 7\right) }$ symmetry at the perturbative
level, as it can be easily realized by considering the dilaton dependence of
loop amplitudes (see \textit{e.g.} \cite{Freedman}). However, this is not
the case for $\mathcal{N}=8$ supergravity. From this perspective, two
(perturbatively finite) theories of quantum gravity would exist, with $32$
local supersymmetries; expectedly, they would differ at least in their
non-perturbative sectors, probed \textit{e.g.} by black hole solutions.
String theorists \cite{GOS,Ark,Banks} claim that $\mathcal{N}=8$, $d=4$
supergravity theory is probably not consistent at the non-perturbative
level. From a purely $d=4$ point of view, their arguments could be overcome
by excluding from the spectrum, as suggested in \cite{BFK-1}, black hole
states which turn out to be singular or ill defined if interpreted as purely
four-dimensional gravitational objects. Inclusion of such singular states
(such as $\frac{1}{4}$-BPS and $\frac{1}{2}$-BPS black holes) would then
open up extra dimensions, with the meaning that a non-perturbative
completion of $\mathcal{N}=8$ supergravity would lead to string theory \cite
{GOS}. Extremal black holes with a consistent $d=4$ interpretation may be
defined as having a Bertotti-Robinson \cite{bertotti} $AdS_{2}\times S^{2}$
near-horizon geometry, with a non-vanishing area of the event horizon. In $%
\mathcal{N}=8$ supergravity, these black holes are\footnote{%
We also remark that these are the only black holes for which the \textit{%
Freudenthal duality} \cite{Duff-Freud-1,FD-1} is well defined.} $\frac{1}{8}$%
-BPS or non-BPS (for a recent review and a list of Refs., see \textit{e.g.}
\cite{ICL-1}). The existence of such states would in any case break the $%
E_{7\left( 7\right) }\left( \mathbb{R}\right) $ continuous symmetry, because
of Dirac-Schwinger-Zwanziger dyonic charge quantization conditions. The
breaking of $E_{7\left( 7\right) }\left( \mathbb{R}\right) $ into an
arithmetic subgroup $E_{7\left( 7\right) }\left( \mathbb{Z}\right) $ would
then manifest only in exponentially suppressed contributions to perturbative
amplitudes (see \textit{e.g.} the discussion in \cite{BHN}, and Refs.
therein), in a similar way to instanton effects in non-Abelian gauge
theories.

The composite anomaly concerns the gauge-scalar sector of the supergravity
theories. Another anomaly, originated in the gravitational part of the
action is the so-called \textit{gravitational anomaly}, which only counts
the basic degrees of freedom associated to the field content of the theory
itself \cite{D-VN,DCGR} (see also \cite{Duff-rev} for a review):
\begin{equation}
g_{\mu \nu }\left\langle T^{\mu \nu }\right\rangle _{1-loop}=\frac{\mathcal{A%
}}{32\pi ^{2}}\int d^{4}x\sqrt{\left| g\right| }\left( R_{\mu \nu \lambda
\rho }^{2}-4R_{\mu \nu }^{2}+R^{2}\right) ,  \label{ga}
\end{equation}
where $\left\langle T^{\mu \nu }\right\rangle _{1-loop}$ is the $1$-loop
\textit{vev} of the gravitational stress-energy tensor. In general, this
trace anomaly is a total derivative and therefore it can be non-vanishing
only on topologically non-trivial $d=4$ backgrounds. Furthermore, as found
long time ago by Faddeev and Popov \cite{FP}, $\left( p+1\right) $-form
gauge fields have a complicated quantization procedure, due to the presence
of ghosts; thus, their contribution to the parameter $\mathcal{A}$ appearing
in the formula (\ref{ga}) vary greatly depending on the field under
consideration. This is because at the quantum level different field
representations are generally inequivalent \cite{D-VN}. Consequently, one
may expect that different formulations of $\mathcal{N}=8$, $d=4$
supergravity (\ref{various-N=8,d=4}), give rise to different gravitational
anomalies. This is actually what happens:

\begin{itemize}
\item  in the formulation $\mathbf{III}$ of (\ref{various-N=8,d=4}) \cite{CJ}%
, with maximal manifest compact symmetry $SO\left( 8\right) $, the
antisymmetric tensors $A_{\mu \nu I}$ are dualized to scalars, and $\mathcal{%
A}\neq 0$.

\item  in the formulation $\mathbf{I}$ of (\ref{various-N=8,d=4}) \cite{CJ},
with maximal manifest compact symmetry $SO\left( 7\right) $, obtained by
compactifying $d=11$ $M$-theory on $T_{7}$, the antisymmetric tensors $%
A_{\mu \nu I}$ are not dualized, and, as found some time ago in \cite{D-VN},
the gravitational anomaly vanishes: $\mathcal{A}=0$. Recently, a wide class
of models has been shown to have $\mathcal{A}=0$, by exploiting \textit{%
generalized mirror symmetry} for seven-manifolds \cite{DF-last-1}.

\item  in the formulation $\mathbf{II}$ of (\ref{various-N=8,d=4}) \cite
{dWN,DCGR} (see also \cite{BEdWN,DP} and the discussion above), with maximal
manifest compact \textit{gauged} symmetry $SO\left( 8\right) $, the
gravitational anomaly is the sum of two contributions: one given by (\ref{ga}%
), and another one related to the non-vanishing cosmological constant $%
\Lambda $, given by
\begin{equation}
\frac{\mathcal{B}}{12\pi ^{2}}\int d^{4}x\sqrt{\left| g\right| }\Lambda ^{2},
\label{ga-Lambda}
\end{equation}
where $\mathcal{B}$, through the relation $\Lambda \sim -e^{2}$ \cite{DCGR},
vanishes whenever the charge $e$ normalization beta function\footnote{$C_{s}$
is the appropriate (positive) \textit{quadratic invariant} for the gauge
group representation in which the particle of spin $s$ sits (see \textit{e.g.%
} Table 1 of \cite{Curtright}, and Refs. therein).} \cite{Curtright}
\begin{equation}
\beta _{e}\left( s\right) =\frac{\hbar }{96\pi ^{2}}e^{3}C_{s}\left(
1-12s^{2}\right) \left( -1\right) ^{2s}
\end{equation}
vanishes, namely in $\mathcal{N}>4$ supergravities (compare \textit{e.g.}
Table II of \cite{DCGR} with Table 1 of \cite{Curtright}). The contribution
to the coefficients $\mathcal{A}$ and $\mathcal{B}$ of (\ref{ga}) and (\ref
{ga-Lambda}) depends on the spin $s$ of the \textit{massless} particle, but
also, as mentioned above, on the its field representation ( \cite{DCGR}; see
also \textit{e.g.} Table 1 of \cite{DF-last-1}):
\begin{eqnarray}
&&
\begin{array}{cccccccc}
\begin{array}{c}
s: \\
~
\end{array}
&
\begin{array}{c}
0~\left( \phi \right) \\
~
\end{array}
&
\begin{array}{c}
0~\left( A_{\mu \nu \rho }\right) \\
~
\end{array}
&
\begin{array}{c}
\frac{1}{2} \\
~
\end{array}
&
\begin{array}{c}
1~\left( A_{\mu }\right) \\
~
\end{array}
&
\begin{array}{c}
1~\left( A_{\mu \nu }\right) \\
~
\end{array}
&
\begin{array}{c}
\frac{3}{2} \\
~
\end{array}
&
\begin{array}{c}
2 \\
~
\end{array}
\\
\begin{array}{c}
360\mathcal{A}: \\
~
\end{array}
&
\begin{array}{c}
4 \\
~
\end{array}
&
\begin{array}{c}
-720 \\
~
\end{array}
&
\begin{array}{c}
7 \\
~
\end{array}
&
\begin{array}{c}
-52 \\
~
\end{array}
&
\begin{array}{c}
364 \\
~
\end{array}
&
\begin{array}{c}
-233 \\
~
\end{array}
&
\begin{array}{c}
848 \\
~
\end{array}
\\
60\mathcal{B}: & -1 & 0 & -3 & -12 & 0 & 137 & -522.
\end{array}
\notag \\
&&
\end{eqnarray}
\end{itemize}

\newpage

\section{Lecture II\newline
(Multi-Center) Black Holes and Attractors}

If $E_{7\left( 7\right) }$ is a continuous non-anomalous symmetry of $%
\mathcal{N}=8$ supergravity, then it is likely that non-perturbative effects
are exponentially suppressed in perturbative amplitudes.

Black holes (BHs) are examples of non-perturbative states which, in presence
of Dirac-Schwinger-Zwanziger dyonic charge quantization, would break $%
E_{7\left( 7\right) }\left( \mathbb{R}\right) $ to a suitable (not unique)
arithmetic subgroup of $E_{7\left( 7\right) }\left( \mathbb{Z}\right) $ (see
\textit{e.g.} \cite{Sen-E7-Z,Duff-Freud-1,BFK-2,ICL-1}, and Refs. therein).

Here we confine ourselves to recall some very basic facts on extremal BHs
(for further detail, see \textit{e.g.} \cite{Haya-rev}, and Refs. therein),
and then we will mention some recent developments on multi-center
solutions.\medskip

For simplicity's sake, we consider the particular class of \textit{extremal}
BH solutions constituted by static, asymptotically flat, spherically
symmetric solitonic objects with dyonic charge vector $\mathcal{Q}$ and
scalars $\phi $ describing trajectories (in the radial evolution parameter $%
r $) with\footnote{%
The subscript ``$H$'' denotes the evaluation at the BH event horizon, whose
radial coordinate is $r_{H}$ (see treatment below).} \textit{fixed points}
determined by the \textit{Attractor Mechanism} \cite{AM-Refs}:

\begin{equation}
\left\{
\begin{array}{l}
\lim_{r\rightarrow r_{H}^{+}}\phi (r)=\phi _{H}(\mathcal{Q}); \\
\\
\lim_{r\rightarrow r_{H}^{+}}\frac{d\phi (r)}{dr}=0.
\end{array}
\right.
\end{equation}
At the horizon, the scalars lose memory of the initial conditions (\textit{%
i.e.} of the asymptotic values $\phi _{\infty }\equiv \lim_{r\rightarrow
\infty }\phi (r)$), and the fixed (attractor) point $\phi _{H}^{a}(\mathcal{Q%
})$ only depends on the BH charges $\mathcal{Q}$. In the supergravity limit,
for $\mathcal{N}>2$ superymmetry, the attractor behavior of such BHs is now
completely classified (see \textit{e.g.} \cite{ADFT-rev,K-rev} for a review
and list of Refs.).

The classical BH entropy is given by the Bekenstein-Hawking entropy-area
formula \cite{BH}

\begin{equation}
S\left( \mathcal{Q}\right) =\frac{A_{H}\left( \mathcal{Q}\right) }{4}=\pi
V_{BH}(\phi _{H}(\mathcal{Q}),\mathcal{Q})=\pi \sqrt{\left| \mathcal{I}%
_{4}\left( \mathcal{Q}\right) \right| }.  \label{sunday0}
\end{equation}
where $V_{BH}$ is the effective BH potential \cite{FGK} (see Eq. (\ref{eff})
below).

The last step of (\ref{sunday0}) holds\footnote{%
Incidentally, the last step of (\ref{sunday0}) also holds for arbitrary
cubic scalar geometry if particular charge configurations are chosen.} for
those theories admitting a quartic polynomial invariant $\mathcal{I}_{4}$ in
the (symplectic) representation of the electric-magnetic duality group in
which $\mathcal{Q}$ sits. This is the case \textit{at least} for the \textit{%
``groups of type }$E_{7}$\textit{''} \cite{Brown}, which are the
electric-magnetic duality groups of supergravity theories in $d=4$ with
\textit{symmetric} scalar manifolds (see \textit{e.g.} \cite{FD-1} for
recent developments, and a list of Refs.). These include all $\mathcal{N}%
\geqslant 3$ supergravities as well as a broad class of $\mathcal{N}=2$
theories in which the vector multiplets' scalar manifold is a \textit{%
special K\"{a}hler} symmetric space (see \textit{e.g.} \cite
{magnific-7,CVP,GST,dWVVP}, and Refs. therein). In the $D$-brane picture of
type $IIA$ supergravity compactified on Calabi-Yau threefolds $CY_{3}$,
charges can be denoted by $q_{0}$ ($D0$), $q_{a}$ ($D2$), $p^{a}$ ($D4$) and
$p^{0}$ ($D6$), and the quartic invariant polynomial $\mathcal{I}_{4}%
\mathcal{\ }$is given by \cite{FG1}
\begin{eqnarray}
\mathcal{I}_{4} &=&-\left( p^{0}q_{0}+p^{a}q_{a}\right) ^{2}+4\left( -p^{0}%
\mathcal{I}_{3}\left( q\right) +q_{0}\mathcal{I}_{3}\left( p\right) +\frac{%
\partial \mathcal{I}_{3}\left( p\right) }{\partial p^{a}}\frac{\partial
\mathcal{I}_{3}\left( q\right) }{\partial q_{a}}\right) ; \\
\mathcal{I}_{3}\left( p\right) &\equiv &\frac{1}{3!}d_{abc}p^{a}p^{b}p^{c};~~%
\mathcal{I}_{3}\left( q\right) \equiv \frac{1}{3!}d^{abc}q_{a}q_{b}q_{c},
\end{eqnarray}
where $d_{abc}$ and $d^{abc}$ are completely symmetric rank-$3$ invariant
tensors of the relevant electric and magnetic charge irreps. of the $U$%
-duality group in $d=5$. A typical (single-center) BPS configuration is $%
\left( q_{0},p^{a}\right) $, with all charges positive (implying $\mathcal{I}%
_{4}>0$), while a typical non-BPS configuration is $\left(
p^{0},q_{0}\right) $ (implying $\mathcal{I}_{4}<0$), see \textit{e.g.} the
discussion in \cite{CFM1} (other charge configurations can be chosen as
well). In the dressed charge basis, manifestly covariant with respect to the
$\mathcal{R}$-symmetry group, the charges arrange into a complex
skew-symmetric central charge matrix $Z_{AB}$. This latter can be
skew-diagonalized to the form \cite{BMZ-Refs}
\begin{equation}
Z_{AB}=\text{diag}\left( z_{1},z_{2},z_{3},z_{4}\right) \otimes \left(
\begin{array}{cc}
0 & -1 \\
1 & 0
\end{array}
\right) ,
\end{equation}
and the quartic invariant can be recast in the following form \cite
{Kallosh-Kol}:
\begin{equation}
\mathcal{I}_{4}=\sum_{i=1}^{4}\left| z_{i}\right|
^{4}-2\sum_{i<j=1}^{4}\left| z_{i}\right| ^{2}\left| z_{j}\right|
^{2}+4\prod_{i=1}^{4}z_{i}+4\prod_{i=1}^{4}\overline{z}_{i}.
\end{equation}
In such a basis, a typical BPS configuration is the one pertaining to the
Reissner-N\"{o}rdstrom BH, with charges $z_{1}=\left( q+ip\right) $ and $%
z_{2}=z_{3}=z_{4}=0$ (implying $\mathcal{I}_{4}=\left( q^{2}+p^{2}\right)
^{2}>0$), whereas a typical non-BPS configuration has (at the event horizon)
$z_{i}=\rho e^{i\pi /4}$ $\forall i=1,...,4$ (implying $\mathcal{I}%
_{4}=-16\rho ^{4}<0$); see \textit{e.g.} the discussion in \cite
{FK-N=8,Gnecchi-2,CFGn-1}.

The simplest example of BH metric is the Schwarzschild BH:

\begin{equation}
ds^{2}=-\,\left( {1-\frac{{2}M}{r}}\right) dt^{2}+\left( {1-\frac{{2}M}{r}}%
\right) ^{-1}dr^{2}+r^{2}d\Omega ^{2},  \label{Schw1}
\end{equation}
where $M$ is the ADM mass \cite{ADM}, and $d\Omega {^{2}=}d\theta {^{2}+}sin{%
^{2}}\theta d\psi {^{2}}$. This BH has no \textit{naked singularity},
\textit{i.e.} the singularity at $r=0$ is \textit{covered} by the event
horizon at $r_{H}=2M$.

The metric (\ref{Schw1}) can be seen as the neutral $q,p\rightarrow 0$ limit
of the Reissner-Nordstr\"{o}m (RN) BH:

\begin{equation}
ds_{RN}^{2}=-\;\left( {1-\frac{{2}M}{r}+\frac{q^{2}+p^{2}}{r{{}^{2}}}}%
\right) dt^{2}+\;\left( {1-\frac{{2}M}{r}+\frac{q^{2}+p^{2}}{r{{}^{2}}}}%
\right) ^{-1}dr^{2}+r^{2}d\Omega ^{2}.  \label{RN-RN-1}
\end{equation}
Such a metric exhibits two horizons, with radii
\begin{equation}
r_{\pm }=M\pm \sqrt{M^{2}-q^{2}-p^{2}}.  \label{sunday-Night-1}
\end{equation}
In the \textit{extremal} case $r_{+}=r_{-}$, and it holds that
\begin{equation}
M^{2}=q^{2}+p^{2},  \label{extr-RN-cond}
\end{equation}
thus a unique event horizon exists at $r_{H}=M$. Notice that for RN BHs the
extremality condition coincides with the saturation of the \textit{BPS bound}
\cite{BPS}
\begin{equation}
M^{2}\geqslant q^{2}+p^{2}.
\end{equation}

By defining $\rho \equiv r-M=r-r_{H}$, the extremal RN metric acquires the
general static Papapetrou-Majumdar \cite{papapetrou} form
\begin{equation}
ds_{RN,extr}^{2}=-\,\left( {1+\frac{M}{\rho }}\right) ^{-2}dt^{2}+\left( {1+%
\frac{M}{\rho }}\right) ^{2}\left( {d\rho ^{2}+\rho ^{2}d\Omega ^{2}}\right)
=-\,e^{2U}dt^{2}+e^{-2U}d\vec{x}^{2},  \label{extr-RN-1}
\end{equation}
where $U=U\left( \overrightarrow{x}\right) $ is an harmonic function
satisfying the $d=3$ Laplace equation
\begin{equation}
\Delta e^{-U\left( \overrightarrow{x}\right) }=0.
\end{equation}

In order to determine the near-horizon geometry of an extremal RN BH, let us
define a new radial coordinate as $\tau =-\frac{1}{\rho }=\frac{1}{r_{H}-r}$%
. Thus, after a further rescaling $\tau \rightarrow \frac{\tau }{M^{2}}$,
the near-horizon limit $\rho \rightarrow 0^{+}$ of extremal metric (\ref
{extr-RN-1}) reads
\begin{equation}
\lim_{\rho \rightarrow 0^{+}}ds_{RN,extr}^{2}=\frac{M^{2}}{\tau ^{2}}\left(
-dt^{2}+{d\tau ^{2}+\tau ^{2}d\Omega ^{2}}\right) ,  \label{BR-BR-1}
\end{equation}
which is nothing but the ${AdS_{2}\times S}^{2}$ Bertotti-Robinson metric
\cite{bertotti}, both \textit{flat} and \textit{conformally flat}.

In presence of scalar fields coupled to the BH background, the BPS bound
gets modified, and in general extremality does not coincide with the
saturation of BPS bound (and thus with supersymmetry preservation) any more.
Roughly speaking, the charges $\mathcal{Q}$ gets ``dressed'' with scalar
fields $\phi $ into the central extension of the local $\mathcal{N}$%
-extended supersymmetry algebra, which is an antisymmetric complex matrix $%
Z_{AB}\left( \phi ,\mathcal{Q}\right) $, named \textit{central charge matrix}
($A,B=1,...,\mathcal{N}$):
\begin{equation}
\left\{
\begin{array}{l}
\left\{ \mathbf{Q}_{\alpha A},\overline{\mathbf{Q}}_{\dot{\alpha}%
}^{B}\right\} =\delta _{A}^{B}\sigma _{\alpha \dot{\alpha}}^{\mu }P_{\mu };
\\
\\
\left\{ \mathbf{Q}_{\alpha A},\mathbf{Q}_{\beta B}\right\} =\epsilon
_{\alpha \beta }Z_{AB}\left( \phi ,\mathcal{Q}\right) .
\end{array}
\right.
\end{equation}
In general
\begin{equation}
Z_{AB}\left( \phi ,\mathcal{Q}\right) =L_{AB}^{\mathbb{A}}\left( \phi
\right) \mathcal{Q}_{\mathbb{A}},
\end{equation}
where $L_{AB}^{\mathbb{A}}\left( \phi \right) $ are the scalar
field-dependent symplectic sections of the corresponding \textit{%
(generalized) special geometry} (see \textit{e.g.} \cite{CDF-rev,FK-N=8,FD-1}%
, and Refs. therein).

In the BH background under consideration, the general \textit{Ans\"{a}tze}
for the vector $2$-form field strengths $F_{\mu \nu }^{\Lambda }$ of the $%
n_{V}$ vector fields ($\Lambda =1,\ldots ,n_{V}$) and their duals $%
G_{\Lambda \mu \nu }=\frac{\delta \mathcal{L}}{\delta F_{\mu \nu }^{\Lambda }%
}$ are given by \cite{FGK}
\begin{eqnarray}
F &=&e^{2U}\mathbb{C}\mathcal{M}(\phi )\mathcal{Q}dt\wedge d\tau +\mathcal{Q}%
\sin {\theta }d\theta \wedge d\psi \,; \\
F &=&\left(
\begin{array}{c}
F_{\mu \nu }^{\Lambda } \\
\\
G_{\Lambda \mu \nu }
\end{array}
\right) \frac{dx^{\mu }dx^{\nu }}{2}\,,
\end{eqnarray}
and electric and magnetic charges $\mathcal{Q}\equiv \left( p^{\Lambda
},q_{\Lambda }\right) ^{T}$ are defined by
\begin{equation}
q_{\Lambda }\equiv \frac{1}{4\pi }\int_{S_{\infty }^{2}}G_{\Lambda
}\,,\qquad p^{\Lambda }\equiv \frac{1}{4\pi }\int_{S_{\infty
}^{2}}F^{\Lambda }\,,
\end{equation}
where $S_{\infty }^{2}$ is the $2$-sphere at infinity. \noindent $\mathcal{M}%
(\phi )$, already discussed in Sec. 1, is a $2n_{V}\times 2n_{V}$ real
symmetric $Sp(2n_{V},\mathbb{R})$ matrix (see Eq. (\ref{M})) whose explicit
form reads \cite{CDF-rev}
\begin{equation}
\mathcal{M}(\phi )=\left(
\begin{array}{cc}
I+RI^{-1}R & -RI^{-1} \\
-I^{-1}R & I^{-1}
\end{array}
\,\right) ,  \label{M-expl}
\end{equation}
with $I\equiv $Im$\,\mathcal{N}_{\Lambda \Sigma }$ and $R\equiv $Re$\,%
\mathcal{N}_{\Lambda \Sigma }$, where $\mathcal{N}_{\Lambda \Sigma }$ is the
(scalar field dependent) kinetic vector matrix entering the $d=4$ Lagrangian
density
\begin{equation}
\mathcal{L}=-\frac{R}{2}+\frac{1}{2}g_{ij}(\phi )\partial _{\mu }\phi
^{i}\partial ^{\mu }\phi ^{j}+I_{\Lambda \Sigma }F^{\Lambda }\wedge ^{\ast
}F^{\Sigma }+R_{\Lambda \Sigma }F^{\Lambda }\wedge F^{\Sigma }\,.
\end{equation}

The black hole effective potential \cite{AM-Refs} is given by
\begin{equation}
V_{BH}\left( \phi ,\mathcal{Q}\right) =-\frac{1}{2}\mathcal{Q}^{T}\mathcal{M}%
\left( \phi \right) \mathcal{Q},  \label{eff}
\end{equation}
This is the effective potential which arises upon reducing the general $%
d\geqslant 4$ Lagrangian on the BH background to the $d=1$ almost geodesic
action describing the radial evolution of the $n_{V}+1$ scalar fields $%
(U(\tau ),\phi ^{i}(\tau ))$ \cite{BreitGM}:
\begin{equation}
S=\int {\mathcal{L}d\tau }=\int (\dot{U}+g_{ij}\dot{\phi}^{i}\dot{\phi}%
^{j}+e^{2U}V_{BH}(\phi (\tau ),p,q)d\tau .
\end{equation}
In order to have the same equations of motion of the original theory, the
action must be complemented with the Hamiltonian constraint, which in the
extremal case reads \cite{FGK}
\begin{equation}
\dot{U}^{2}+g_{ij}\dot{\phi}^{i}\dot{\phi}^{j}-e^{2U}V_{BH}(\phi (\tau
),p,q)=0\,.
\end{equation}

The black hole effective potential $V_{BH}$ can generally be written in
terms of the superpotential $W(\phi )$ as
\begin{equation}
V_{BH}=W^{2}+2g^{i{j}}\partial _{i}W\partial _{j}W\,\,.  \label{VW}
\end{equation}
This formula can be viewed as a differential equation defining $W$ for a
given $V_{BH}$, and it can lead to multiple choices, one corresponding to
BPS solutions, and the others associated to non-BPS ones. $W$ allows to
rewrite the ordinary second order supergravity equations of motion
\begin{eqnarray}
\ddot{U} &=&e^{2U}V_{BH}; \\
\ddot{\phi}^{i} &=&g^{ij}\frac{\partial V_{BH}}{\partial \phi _{j}}e^{2U}\,,
\end{eqnarray}
as first order flow equations, defining the radial evolution of the scalar
fields $\phi ^{i}$ and the warp factor $U$ from asymptotic (radial) infinity
towards the black hole horizon \cite{Ceresole:2007wx} :
\begin{equation}
\dot{U}=-e^{U}W\,,\qquad \qquad \dot{\phi}^{i}=-2e^{U}g^{i{j}}\partial
_{j}W\,.  \label{1.18}
\end{equation}
\noindent At the prize of finding a suitable \textit{``fake''} first order
superpotential $W$, one only has to deal with these first order flow
equations even for non-supersymmetric solutions, where one does not have
Killing spinor equations \cite{Ceresole:2007wx,Andrianopoli:2007gt}.

For $\frac{1}{\mathcal{N}}$-BPS supersymmetric BHs in $\mathcal{N}\geqslant 2
$ supergravity theories (with central charge matrix $Z_{AB}$), $\mathcal{W}$
is given by the square root\footnote{%
The subscript ``$h$'' stands for ``the highest''.} $\sqrt{\lambda _{h}}$ of
the largest of the eigenvalues of $Z_{AB}Z^{\dag BC}$ \cite
{Ceresole:2007wx,Andrianopoli:2007gt}. Furthermore, $\mathcal{W}$ has a
known analytical expression for all $\mathcal{N}\geqslant 2$ charge
configurations with $\mathcal{I}_{4}>0$ (for $\mathcal{N}=2$, this applies
to special K\"{a}hler geometry based on symmetric spaces, see \textit{e.g.}
\cite{dWVVP}) \cite{Andrianopoli:2007gt}. For $\mathcal{I}_{4}<0$, $\mathcal{%
W}^{2}$ has an analytical expression for rank-$1$ and rank-$2$ cosets \cite
{CDFY-1,BMP-1,CDFY-2}, while it is known to exist in general as a solution
of a sixth order algebraic equation \cite{BMP-1,CDFY-2,FMO-1}.

The Bekenstein-Hawking BH entropy \cite{BH} (\ref{sunday0}) can be written
in terms of $W$ as follows:
\begin{equation}
S\left( \mathcal{Q}\right) =\pi \left. W^{2}\right| _{\partial W=0},
\end{equation}
where the critical points of the suitable $W$ reproduce a class of critical
points of $V$ itself. It is worth remarking that the value of the
superpotential $W$ at radial infinity also encodes other basic properties of
the extremal black hole, namely its $ADM$ mass \cite{ADM}, given by ($\phi
_{\infty }^{i}\equiv \lim_{r\rightarrow \infty }\phi ^{i}\left( r\right) $)
\begin{equation}
M_{ADM}(\phi _{\infty },\mathcal{Q})=\dot{U}(\tau =0)=W(\phi _{\infty },%
\mathcal{Q}),
\end{equation}
and the scalar charges
\begin{equation}
\Sigma ^{i}\left( \phi _{\infty },\mathcal{Q}\right) =2g^{ij}(\phi _{\infty
})\frac{\partial W}{\partial \phi ^{i}}(\phi _{\infty },\mathcal{Q}).
\end{equation}
\bigskip

\textit{Multi-center} BHs are a natural extension of single-center BHs, and
they play an important role in the dynamics of quantum theories of gravity,
such as superstrings and $M$-theory.

In fact, interesting multi-center solutions have been found for BPS BHs in $%
d=4$ theories with $\mathcal{N}=2$ supersymmetry, in which the \textit{%
Attractor Mechanism} \cite{AM-Refs,FGK} is generalized by the so-called
\textit{split attractor flow} \cite{D-1}. This name comes from the
existence, for $2$-center solutions, of a co-dimension one region (named
\textit{marginal stability (MS) wall)} in the scalar manifold, where in fact
a stable $2$-center BH configuration may decay into two single-center
constituents, whose scalar flows then separately evolve according to the
corresponding attractor dynamics.

The study of these phenomena has recently progressed in many directions. By
combining properties of $\mathcal{N}=2$ supergravity and superstring theory,
a number of interesting phenomena, such as split flow tree, entropy enigma,
bound state recombination walls, and microstate counting have been
investigated (see \textit{e.g. }\cite{BD-1,Split-Refs,WC-2,MS-FM-1}).

The MS wall is defined by the condition of stability for a marginal decay of
a $2$-center BH compound solution with charge $\mathcal{Q}=\mathcal{Q}_{1}+%
\mathcal{Q}_{2}$ into two single-center BHs (respectively with charges $%
\mathcal{Q}_{1}$ and $\mathcal{Q}_{2}$):
\begin{equation}
M\left( \phi _{\infty },\mathcal{Q}_{1}+\mathcal{Q}_{2}\right) =M\left( \phi
_{\infty },\mathcal{Q}_{1}\right) +M\left( \phi _{\infty },\mathcal{Q}%
_{2}\right) .  \label{MS-cond}
\end{equation}
As mentioned, after crossing the MS wall each flow evolves towards its
corresponding attractor point, and the classical entropy of each BH
constituent follows the Bekenstein-Hawking formula (\ref{sunday0}). It
should be noted that the entropy of the original compound (conceived as a
\textit{single-center} BH with total charge $\mathcal{Q}=\mathcal{Q}_{1}+%
\mathcal{Q}_{2}$) can be smaller, equal, or larger than the sum of the
entropies of its constituents:
\begin{equation}
S\left( \mathcal{Q}_{1}+\mathcal{Q}_{2}\right) \gtreqless S\left( \mathcal{Q}%
_{1}\right) +S\left( \mathcal{Q}_{2}\right) .
\end{equation}

For $\mathcal{N}=2$ BPS compound and constituents in $\mathcal{N}=2$, $d=4$
supergravity (in which $Z_{AB}=\epsilon _{AB}Z$), (\ref{MS-cond}) can be
recast as a condition on the central charge ($Z_{i}\equiv M\left( \phi
_{\infty },\mathcal{Q}_{i}\right) $, $i=1,2$, and $Z_{1+2}\equiv Z\left(
\phi _{\infty },\mathcal{Q}_{1}+\mathcal{Q}_{2}\right) =Z_{1}+Z_{2}$):
\begin{equation}
\left| Z_{1}+Z_{2}\right| =\left| Z_{1}\right| +\left| Z_{2}\right| .
\label{MS-1}
\end{equation}
Furthermore, before crossing the MS wall, the relative distance $\left|
\overrightarrow{x_{1}}-\overrightarrow{x_{2}}\right| $ of the two BH
constituents with \textit{mutually non-local }charges $\left\langle \mathcal{%
Q}_{1},\mathcal{Q}_{2}\right\rangle \neq 0$ is given by \cite{BD-1}
\begin{equation}
\left| \overrightarrow{x_{1}}-\overrightarrow{x_{2}}\right| =\frac{1}{2}%
\frac{\left\langle \mathcal{Q}_{1},\mathcal{Q}_{2}\right\rangle \left|
Z_{1}+Z_{2}\right| }{\text{Im}\left( Z_{1}\overline{Z_{2}}\right) },
\label{N=2}
\end{equation}
where
\begin{equation}
2\left| \text{Im}\left( Z_{1}\overline{Z_{2}}\right) \right| =\sqrt{4\left|
Z_{1}\right| ^{2}\left| Z_{2}\right| ^{2}-\left( \left| Z_{1}+Z_{2}\right|
^{2}-\left| Z_{1}\right| ^{2}-\left| Z_{2}\right| ^{2}\right) ^{2}}.
\label{N=2--}
\end{equation}
Correspondingly, the 2-center BH has an intrinsic (orbital) angular
momentum, given by \cite{BD-1}
\begin{equation}
\overrightarrow{J}=\frac{1}{2}\left\langle \mathcal{Q}_{1},\mathcal{Q}%
_{2}\right\rangle \frac{\overrightarrow{x_{1}}-\overrightarrow{x_{2}}}{%
\left| \overrightarrow{x_{1}}-\overrightarrow{x_{2}}\right| }.
\end{equation}

Note that when the charge vectors $\mathcal{Q}_{1}$ and $\mathcal{Q}_{2}$
are \textit{mutually local} \ (\textit{i.e.} $\left\langle \mathcal{Q}_{1},%
\mathcal{Q}_{2}\right\rangle =0$), $\left| \overrightarrow{x_{1}}-%
\overrightarrow{x_{2}}\right| $ is not constrained at all, and $J=0$.
Actually, this is always the case for the scalarless case of extremal
Reissner-N\"{o}rdstrom double-center BH solutions in $\mathcal{N}=2$ \textit{%
pure} supergravity. Indeed, in this case the central charge simply reads
(see also discussion above)
\begin{equation}
Z_{RN}\left( p,q\right) =q+ip,
\end{equation}
and it is immediate to check that the marginal stability condition (\ref
{MS-1}) implies $\left\langle \mathcal{Q}_{1},\mathcal{Q}_{2}\right\rangle
=q_{1}p_{2}-p_{1}q_{2}=0$.

It is here worth observing that Im$\left( Z_{1}\overline{Z_{2}}\right) =0$
both describes marginal and \textit{anti-marginal} stability \cite{WC-2}.
\textit{Marginal stability} further requires
\begin{equation}
\text{Re}\left( Z_{1}\overline{Z_{2}}\right) >0\Leftrightarrow \left|
Z_{1}+Z_{2}\right| ^{2}>\left| Z_{1}\right| ^{2}+\left| Z_{2}\right| ^{2}.
\label{MS-branch}
\end{equation}
The other (unphysical) branch, namely
\begin{equation}
\text{Re}\left( Z_{1}\overline{Z_{2}}\right) <0\Leftrightarrow \left|
Z_{1}+Z_{2}\right| ^{2}<\left| Z_{1}\right| ^{2}+\left| Z_{2}\right| ^{2},
\label{AMS-branch}
\end{equation}
pertains to \textit{anti-marginal stability}, reached for $\left|
Z_{1}+Z_{2}\right| =\left| \left| Z_{1}\right| -\left| Z_{2}\right| \right| $%
.

Eq. (\ref{N=2}) implies the stability region for the $2$-center BH solution
to occur for
\begin{equation}
\left\langle \mathcal{Q}_{1},\mathcal{Q}_{2}\right\rangle \text{Im}\left(
Z_{1}\overline{Z_{2}}\right) >0,
\end{equation}
while it is forbidden for $\left\langle \mathcal{Q}_{1},\mathcal{Q}%
_{2}\right\rangle $Im$\left( Z_{1}\overline{Z_{2}}\right) <0$. The scalar
flow is directed from the stability region towards the instability region,
crossing the MS wall at $\left\langle \mathcal{Q}_{1},\mathcal{Q}%
_{2}\right\rangle $Im$\left( Z_{1}\overline{Z_{2}}\right) =0$. This implies
that the stability region is placed \textit{beyond} the MS wall, and \textit{%
on the opposite side} of the split attractor flows.

By using the fundamental identities of $\mathcal{N}=2$ special K\"{a}hler
geometry in presence of two (mutually non-local) symplectic charge vectors $%
\mathcal{Q}_{1}$ and $\mathcal{Q}_{2}$ (see \textit{e.g.} \cite
{D-1,BFM-1,FK-N=8}), one can compute that at BPS attractor points of the
centers $1$ \textit{or} $2$:
\begin{equation}
\left\langle \mathcal{Q}_{1},\mathcal{Q}_{2}\right\rangle =-2\text{Im}\left(
Z_{1}\overline{Z_{2}}\right) \Rightarrow 2\left\langle \mathcal{Q}_{1},%
\mathcal{Q}_{2}\right\rangle \text{Im}\left( Z_{1}\overline{Z_{2}}\right)
=-\left\langle \mathcal{Q}_{1},\mathcal{Q}_{2}\right\rangle ^{2}<0.
\label{N=2-1}
\end{equation}
By using (\ref{N=2}) and (\ref{N=2-1}), one obtains $\left| \overrightarrow{%
x_{1}}-\overrightarrow{x_{2}}\right| <0$: this means that, as expected, the
BPS attractor points of the centers $1$ \textit{or} $2$ do not belong to the
stability region of the $2$-center BH solution. Furthermore, the result (\ref
{N=2-1}) also consistently implies:
\begin{eqnarray}
&&
\begin{array}{l}
\text{stability region}: \\
\left\langle \mathcal{Q}_{1},\mathcal{Q}_{2}\right\rangle \text{Im}\left(
Z_{1}\overline{Z_{2}}\right) =\left| \left\langle \mathcal{Q}_{1},\mathcal{Q}%
_{2}\right\rangle \right| \sqrt{4\left| Z_{1}\right| ^{2}\left| Z_{2}\right|
^{2}-\left( \left| Z_{1}+Z_{2}\right| ^{2}-\left| Z_{1}\right| ^{2}-\left|
Z_{2}\right| ^{2}\right) ^{2}}>0;
\end{array}
\notag \\
&&\text{~}  \label{stab-region} \\
&&
\begin{array}{l}
\text{instability region}: \\
\left\langle \mathcal{Q}_{1},\mathcal{Q}_{2}\right\rangle \text{Im}\left(
Z_{1}\overline{Z_{2}}\right) =-\left| \left\langle \mathcal{Q}_{1},\mathcal{Q%
}_{2}\right\rangle \right| \sqrt{4\left| Z_{1}\right| ^{2}\left|
Z_{2}\right| ^{2}-\left( \left| Z_{1}+Z_{2}\right| ^{2}-\left| Z_{1}\right|
^{2}-\left| Z_{2}\right| ^{2}\right) ^{2}}<0,
\end{array}
\notag \\
&&  \label{split-flow-region}
\end{eqnarray}
where a particular case of (\ref{split-flow-region}), holding at the
attractor points, is given by (\ref{N=2-1}).\smallskip

As shown in \cite{MS-FM-1}, by exploiting the theory of \textit{matrix norms}%
, all above results can be extended \textit{at least} to $\mathcal{N}=2$
non-BPS states with $\mathcal{I}_{4}>0$, as well as to BPS states in $%
\mathcal{N}>2$ supergravity.

For two-center BHs, by replacing $\left| Z\right| $ with$\sqrt{\lambda _{h}}$%
, the generalization of (\ref{N=2}) \textit{e.g.} to $\mathcal{N}=8$ maximal
supergravity reads
\begin{equation}
\left| \overrightarrow{x_{1}}-\overrightarrow{x_{2}}\right| =\frac{\left|
\left\langle \mathcal{Q}_{1},\mathcal{Q}_{2}\right\rangle \right| \sqrt{%
\lambda _{1+2,h}}}{\sqrt{4\lambda _{1,h}\lambda _{2,h}-\left( \lambda
_{1+2,h}-\lambda _{1,h}-\lambda _{2,h}\right) ^{2}}},  \label{N=8-gen}
\end{equation}
where $\lambda _{1+2,h}\equiv \lambda _{h}\left( \phi _{\infty },\mathcal{Q}%
_{1}+\mathcal{Q}_{2}\right) $ and $\lambda _{i,h}\equiv \lambda _{h}\left(
\phi _{\infty },\mathcal{Q}_{i}\right) $.

Analogously, also result (\ref{N=2-1}) can be generalized \textit{e.g.} to
suitable states in $\mathcal{N}=8$ supergravity. Indeed, by exploiting the $%
\mathcal{N}=8$ generalized special geometry identities \cite{FK-N=8} ($%
\mathbf{Z}_{i}\equiv Z_{AB}\left( \phi _{\infty },\mathcal{Q}_{i}\right) $)
\begin{equation}
\left\langle \mathcal{Q}_{1},\mathcal{Q}_{2}\right\rangle =-\text{Im}\left(
Tr\left( \mathbf{Z}_{1}\mathbf{Z}_{2}^{\dag }\right) \right) ,
\label{N=8-Ids}
\end{equation}
one can compute that at the $\frac{1}{8}$-BPS attractor points of the
centers $1$ \textit{or} $2$ it holds
\begin{equation}
\left| \left\langle \mathcal{Q}_{1},\mathcal{Q}_{2}\right\rangle \right| =%
\sqrt{4\lambda _{h,1}\lambda _{h,2}-\left( \lambda _{1,h}+\lambda
_{2,h}-\lambda _{1+2,h}\right) ^{2}}.  \label{N=8-gen-1}
\end{equation}
Analogously to the $\mathcal{N}=2$ case treated above, note that $\frac{1}{8}
$-BPS attractor points of the centers $1$ \textit{or} $2$ do not belong to
the stability region of the two-center BH solution, but instead they are
placed, with respect to the stability region, on the opposite side of the MS
wall.

\section*{Acknowledgments}

The work of S. F. is supported by the ERC Advanced Grant no. 226455, \textit{%
``Supersymmetry, Quantum Gravity and Gauge Fields''}
(\textit{SUPERFIELDS}).

\end{document}